\documentclass[12pt,preprint]{aastex}
\usepackage{graphicx}

\def\be{\begin{equation}}
\def\ee{\end{equation}}
\def\ba{\begin{eqnarray}}
\def\ea{\end{eqnarray}}
\newcommand{\msun}{\ifmmode\mbox{M}_{\odot}\else$\mbox{M}_{\odot}$\fi}
\newcommand{\rsun}{\ifmmode\mbox{R}_{\odot}\else$\mbox{M}_{\odot}$\fi}
\newcommand{\degrees}{\ifmmode^{\circ}\else$^{\circ}$\fi}
\newcommand{\degree}{\ifmmode^{\circ}\else$^{\circ}$\fi}
\newcommand{\amin}{\ifmmode^{\prime}\else$^{\prime}$\fi}
\newcommand{\asec}{\ifmmode^{\prime\prime}\else$^{\prime\prime}$\fi}

\shorttitle{AO327, an Arecibo 327~MHz Drift Pulsar Survey}
\shortauthors{Deneva et al.}

\begin{document}

\title{Goals, Strategies and First Discoveries of AO327, the Arecibo All-Sky 327~MHz Drift Pulsar Survey} 
\author{
J.~S.~Deneva$^{1,*}$, 
K.~Stovall$^{2,3}$,
M.~A.~McLaughlin$^{4}$,
S.~D.~Bates$^{4}$,
P.~C.~C.~Freire$^{5}$,
J.~G.~Martinez$^{2}$,
F.~Jenet$^{2}$,
M.~Bagchi$^{4}$}
\affil{$^{1}$Arecibo Observatory, HC3 Box 53995, Arecibo, PR 00612}
\affil{$^{2}$Center for Advanced Radio Astronomy, Department of Physics and Astronomy, University of Texas at Brownsville, Brownsville, TX 78520}
\affil{$^{3}$Department of Physics and Astronomy, University of Texas at San Antonio, San Antonio, TX 78249}
\affil{$^{4}$Department of Physics, West Virginia University, 111 White Hall, Morgantown, WV 26506}
\affil{$^{5}$Max-Planck-Institut f\"{u}r Radioastronomie, Auf dem H\"{u}gel 69, D-53121 Bonn, Germany}

\begin{abstract}
We report initial results from AO327, a drift survey for pulsars with the Arecibo telescope at 327~MHz. The first phase of AO327 will cover the sky at declinations of $-1\degrees$ to $28\degrees$, excluding the region within $5\degrees$ of the Galactic plane, where high scattering and dispersion make low-frequency surveys sub-optimal. We record data from a 57~MHz bandwidth with 1024 channels and 125~$\mu$s sampling time. The 60~s transit time through the AO327 beam means that the survey is sensitive to very tight relativistic binaries even with no acceleration searches. To date we have detected 44 known pulsars with periods ranging from 3~ms to 2.21~s and discovered 24 new pulsars. The new discoveries include three millisecond pulsars, three objects with periods of a few tens of milliseconds typical of young as well as mildly recycled pulsars, a nuller, and a rotating radio transient. Five of the new discoveries are in binary systems. The second phase of AO327 will cover the sky at declinations of $28\degrees - 38\degrees$. We compare the sensitivity and search volume of AO327 to the Green Bank North Celestial Cap survey and the GBT350 drift survey, both of which operate at 350~MHz. %{\bf [XXX update number and period range for redetections and new pulsars before submission]}

\end{abstract}

\section{Introduction}

Currently there are $\sim 2000$ known pulsars, the majority in our Galaxy and a few in the Magellanic Clouds. About 250 pulsars have rotation periods of less than ten milliseconds and are nature's most accurate clocks. The unique properties of these millisecond pulsars (MSPs) can be exploited to study the exotic state of matter in the interiors of neutron stars, to detect gravitational waves, and to test the theory of general relativity in the strong gravity regime. For detecting gravitational waves using a Pulsar Timing Array (PTA; e.g. \citealt{Jenet05}), one needs regular observations of MSPs that are extremely stable rotators and are, optimally, distributed all over the sky. 

Most pulsars have been discovered in large radio surveys, but past survey coverage has favored the southern hemisphere and the region within a few degrees of the Galactic plane. The Arecibo drift-scan surveys carried out at 430 MHz prior to and during the Gregorian dome upgrade in the 1990s covered about 70\% of the Arecibo sky (declinations between $-0.5\degrees$ and 38\degrees, \citealt{xs98}). Among the many pulsars found, some have been extremely important for science in general and radio astronomy in particular. These include: PSR~B1534+12 \citep{sac+98}, the relativistic binary system which has allowed two different tests of general relativity; PSR~B1257+12 \citep{wf92}, a MSP now known to be orbited by three planets and the first extra-solar planetary system discovered; and PSR~J1713+0747 \citep{fwc93}, a bright, stable MSP which dominates the sensitivity of the PTA \citep{Demorest13}. 
%The other two most stable rotators known in addition to PSR~J1713+0747 are PSR~J1909$-$3744 and PSR~J0437$-$4715. They are also part of the PTA and were found in high-latitude surveys with the Parkes telescope (\citealt{Johnston93}, \citealt{Jacoby05}).

Despite the successes of the Arecibo searches made so far, they also had limitations. The sky was divided into five regions, with each of these observed by a different group. Early observations were made with old equipment, which had coarser time and spectral resolution than is available from current instruments. Data collected during the upgrade were often of variable quality and the different groups  were using different search algorithms and methods for interference excision. As a result, widely different success rates in finding new pulsars were reported. This complicates any efforts to draw inferences about the general pulsar population probed by these surveys. Given the patchy nature of the observations and data processing carried out so far, there is good reason to believe that $\sim 30\%$ of the Arecibo sky remains to be surveyed, and that many pulsars await discovery \citep{xs98}. This belief is supported by the discovery of 11 new pulsars in an analysis of 430~MHz drift data covering 490 deg$^2$ by \cite{lma+04} and \cite{lxf+05}. Two of the new pulsars have periods $<60$ ms. 

For a few months in 2007, the Green Bank Telescope (GBT) underwent azimuth track repair, which made normal pointed observations impossible, but \cite{Lynch13} and \cite{Boyles13} used $\sim 1500$~h to conduct a 350~MHz drift survey (GBT350) of sky in the declination range between $-21\degrees$ and 2\degrees\ (which is mostly inaccessible to the Arecibo telescope) and to a lesser extent at declinations of $17\degrees - 26$\degrees. GBT350 has discovered 35 pulsars, including 7 MSPs. One of the new MSPs is in a hierarchical triple system \citep{Ransom13}, one of only two such systems known. Another MSP, PSR~J1023+0038, was the first object whose evolution from a low-mass X-ray binary to a radio-emitting MSP was observed almost in real time over the span of a decade \citep{Archibald09}. A third discovery, the mildly recycled pulsar J0348$+$0432, is in a binary system with an orbital period of only 2.4~h, and has the highest precisely measured NS mass, $2.01 \pm 0.04$~\msun \citep{Antoniadis13}. This makes it a unique lab for testing alternative theories of gravity, because it probes the properties of gravity for the most extreme gravitational fields. Since 2009, the Green Bank north celestial cap survey (GBNCC, \citealt{GBNCC}) has searched the sky at declinations $> 38\degrees$ (also inaccessible to the Arecibo telescope) at 350~MHz and has discovered 62 new pulsars, including 9 MSPs. The results of GBT350 and GBNCC demonstrate the excellent scientific potential of high-latitude pulsar surveys.

In Section~\ref{sec_survey} we discuss the current status of AO327 in terms of sky coverage, sensitivity, and search volume, and we make comparisons with the Green Bank low-frequency surveys. Section~\ref{sec_new} focuses on the 24 new pulsars discovered by AO327 so far, and Section~\ref{sec_future} presents our plans and projections for the next few years that will be necessary to survey the entire Arecibo sky at 327~MHz.

\section{The AO327 Survey}\label{sec_survey}

The 1-minute effective integration time of an Arecibo drift survey at 327~MHz means that it is sensitive to very tight relativistic binaries even with only moderate acceleration searches. The double pulsar J0737$-$3039 has an orbital period of 2.46~h and was found in a Parkes survey with an integration time of 265~s without using an acceleration search during data processing \citep{Burgay06}. This means that for similar flux densities, companion masses, and spin periods, AO327 is sensitive to comparable binaries with orbital periods of $2.46~{\rm h} \times 60/265 = 0.57$~h with a conventional periodicity search, and to even shorter-period comparable binaries if using acceleration search. Pulsars in tight relativistic binaries with orbital periods of a few hours or less can be timed very well at Arecibo, where the maximum source tracking time is $\sim 2.5 - 3$~h. 

\begin{figure}[t]
\centering
\includegraphics[width=\textwidth]{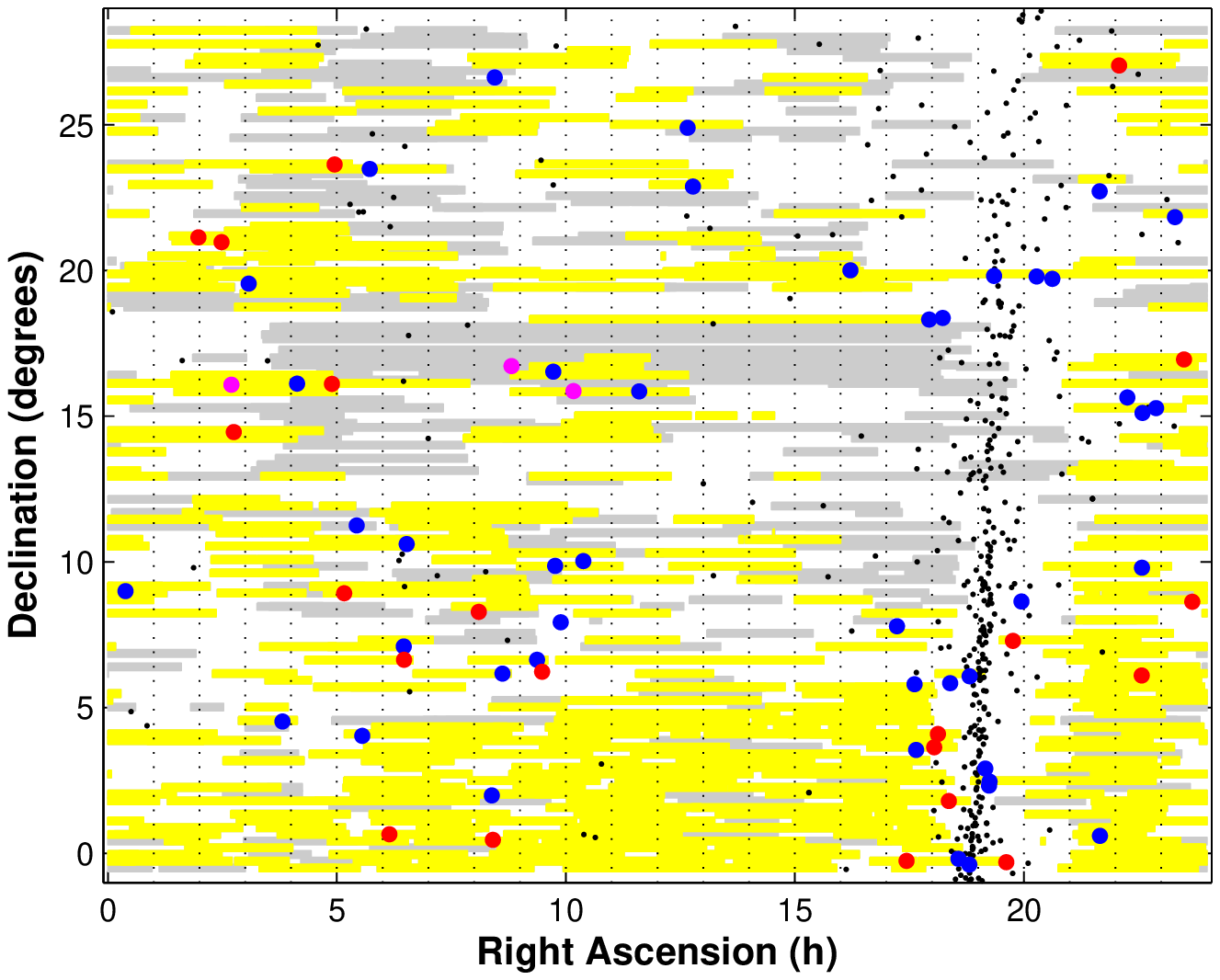}
\caption{Current sky coverage of the AO327 survey. Grey strips indicate data taken. Yellow strips indicate data processed and searched. Black dots correspond to known pulsars from the ATNF catalog. Blue circles show known pulsars detected blindly by the AO327 search pipeline. Red circles show new pulsars discovered by AO327, and magenta circles show new pulsars discovered during the WAPP-based pilot 327~MHz drift survey from 2003. \label{fig_coverage}}
\end{figure}

We conducted a pilot 327~MHz drift survey at Arecibo in 2003 for 100~h. The frequency, lower than used in previous Arecibo drift surveys, was chosen because pulsars tend to have power-law spectra that decline steeply with increasing frequency ($S \propto \nu^{\alpha}$, with average $\alpha = -1.7$). It also anticipated the installation of a new, cooled 327~MHz receiver, which was eventually commissioned in 2006. This pilot survey used the older Wideband Arecibo Pulsar Processor (WAPP, \citealt{Dowd00}) backend and discovered three pulsars (Table~\ref{tab_new}, Figure~\ref{fig_coverage}). 

With the cooled 327~MHz receiver and the Mock spectrometer\footnote{\tt http://www.naic.edu/\~{}phil/hardware/pdev/pdev.html} available, we began taking data during a lengthy repair of the telescope platform in 2010, which restricted telescope movement for several months. The Mock spectrometer allows us to collect data from a 57~MHz bandwidth as opposed to 25~MHz for the WAPP-based pilot survey and has a better response to radio frequency interference (RFI) than the older backend. 

In 2012, the Arecibo Observatory acquired a new backend, the Puerto Rico Ultimate Pulsar Processing Instrument (PUPPI), which is a clone of the similarly named GUPPI at Green Bank\footnote{\tt http://safe.nrao.edu/wiki/bin/view/CICADA/GUPPISupportGuide}. PUPPI provides a ``fast4k'' search mode, allowing recording of 4096 channels with a minimum sampling time of 81.92~$\mu$s. After PUPPI control is integrated into the Arecibo telescope observing software, AO327 will begin using PUPPI for regular survey observations. In the meantime, we have successfully used PUPPI in this mode to confirm some of the new pulsars in Table~\ref{tab_new}, and in coherent dedispersion mode we will use it for timing AO327 discoveries. Because of hardware constraints, the smallest bandwidth available with PUPPI is 100~MHz, while the 327~MHz receiver bandwidth is 67~MHz.  A modification in the PUPPI datataking software will allow us to record data only from the 68~MHz receiver bandwidth although the 4096 channels of the ``fast4k'' mode cover 100~MHz. This accounts for the 2816 channels listed in Table~\ref{tab_surveys} for the future AO327 PUPPI setup. 

% Verify GBT350 and GBNCC parameters from papers
\begin{table}
\caption{Low-frequency survey parameters used in comparison.\label{tab_surveys}}
\begin{center}
\begin{tabular}{lccccc}
\hline
Parameter & AO327 Pilot & AO327 & AO327 & GBT350 & GBNCC \\
\hline
Backend & WAPP & Mock & PUPPI & Spigot & GUPPI \\
$f_{\rm center}$ (MHz) & 327 & 327 & 327 & 350 & 350 \\
$T_{\rm rec}$ (K) & 130 & 113 & 113 & 46 & 46 \\
$G$ (K/Jy) & 11 & 11 & 11 & 2 & 2 \\
$T_{\rm obs}$ (s) & 60 & 60 & 60 & 140 & 120 \\
$\Delta\nu$ (MHz) & 25 & 57 & 69 & 50 & 100 \\
$N_{\rm ch}$  & 512 & 1024 & 2816 & 2048 & 4096 \\
$\Delta t$ ($\mu$s) & 256 & 125 & 82 & 82 & 82 \\
\hline
\end{tabular}
\end{center}
\end{table}

\subsection{Coverage}

The AO327 survey aims to search the entire Arecibo sky (declinations between $-1\degrees$ and 38\degrees) for pulsars and radio transients at 327~MHz. We plan to do this in two phases. During the 2010 platform repair, telescope movement was restricted successively to several different azimuth and zenith angle ranges, giving access to declinations (DEC) of $0 - 28$\degrees. The necessity to coordinate observations with repair work and regularly scheduled projects that were able to use the telescope in this restricted mode resulted in patchy sky coverage. After the movement restrictions were lifted in early 2011, AO327 continued to observe during time slots that were unassigned or became available at short notice (e.g. due to radar transmitter malfunction) until a proposal for dedicated telescope time was submitted and accepted. 

Our Phase 1 goal is to fill in coverage gaps in the DEC~$= 0 - 28$\degrees\ range (Figure~\ref{fig_coverage}), excluding the region within $\pm 5$ deg of the inner Galactic plane. The PALFA survey \citep{PALFA} already covers the latter region at 1.4~GHz, a frequency better suited for pulsar searching in the high dispersion and scattering environment of the Galactic plane. We also include in Phase 1 the remaining one-degree strip at the south edge of sky accessible to Arecibo (DEC~$= -1 - 0$\degrees) that was not covered during the period of platform repair. Phase 2 will extend the survey to the remaining portion of the Arecibo sky,  DEC~$= 28 - 38$\degrees. So far the AO327 has taken survey data with the Mock spectrometer for a total of 1687~h. In order to completely cover the field shown in Figure~\ref{fig_coverage}, a total of 2707~h are necessary. So far we have 1413~h of non-redundant sky coverage and therefore at least 1294~h are necessary to fill in all gaps in the Phase 1 survey region. AO327 gets $\sim 400$~h of telescope time per year. However, this includes setup and slewing overhead in addition to actual data-taking time. In addition, some observing time is used for candidate confirmations. We estimate that Phase 1 will be completed by 2017. Phase 2 of the survey will take two additional years. 

The 274~h of data corresponding to redundant sky coverage are due to limited telescope mobility in 2010 and the combined partial overlap in right ascension (RA) of adjacent drift strips. In addition, the 15\amin\ strip of sky centered at DEC~$= 19^{\rm h}52\amin$) has been observed for $2.5 \times 24$~h. This DEC corresponds to the telescope stow position. During observatory shut-down due to tropical storm or hurricane warnings as well as in actual severe tropical weather conditions, the telescope is locked in the stow position, but it is still possible to take data in drift mode. AO327 observes under such conditions and we have detected several known pulsars in data from this ``hurricane strip''. Redundant sky coverage improves the chance of detecting transient, repeating sources, scintillating, eclipsing, and intermittent pulsars, and weak or long-period pulsars whose detectability in a blind search may be affected by variable RFI. Case in point is PSR~J0241+16 (Table~\ref{tab_new}), which was discovered in the pilot WAPP-based 327~MHz drift search from 2003 and has been confirmed in a longer observation, but was not detected via a blind search in Mock AO327 data covering the same strip of sky. Folding the corresponding data span with the pulsar's period ($P$) and dispersion measure (DM) produced a marginal detection contaminated by RFI. This pulsar also has the second lowest DM among our discoveries, which makes it especially susceptible to scintillation.

\subsection{Sensitivity}

We calculate the mininum detectable pulsar flux density for a survey using the equation
\be
S_{\rm min} = \frac{\beta\left(S/N\right)_{\rm min} T_{\rm sys}}{G \sqrt{N_{\rm p} T_{\rm obs} \Delta\nu}} \sqrt{\frac{W}{P - W}},
\ee
where $\left(S/N\right)_{\rm min} = 6$ is the minimum signal-to-noise ratio corresponding to a detection, $T_{\rm sys} = T_{\rm rec} + T_{\rm sky}$ is a system temperature consisting of receiver and sky terms, $G$ is the gain, $N_{\rm p} = 2$ is the number of polarization channels summed, $T_{\rm obs}$ is the maximum transit time through the beam, $\Delta\nu$ is the bandwidth, $W$ is the observed pulse width, and $P$ is the pulsar period. The observed pulse width $W = \sqrt{W_{\rm int}^2 + \Delta t_{\rm DM, 1ch}^2 + \Delta t_{\rm \delta DM}^2 + \Delta t_{\rm samp} ^2+ \tau_{\rm s}^2}$. For our calculations, we use $T_{\rm sky} = 50$~K \citep{Haslam82} and assume an intrinsic pulse width $W_{\rm int} = 0.1~P$. The factor $\beta$ expresses sensitivity degradation due to digitization effects. It is unity for the Mock spectrometer, PUPPI, and GUPPI, and $\approx 1.16$ for 3-level correlators like the Spigot and WAPP. 

To calculate the dispersion smearing over one channel ($\Delta t_{\rm DM, 1ch}$) and the dispersion smearing due to DM error in the worst case, when the pulsar DM is halfway between two trial DMs ($\Delta t_{\rm \delta DM}$), we use the cold plasma dispersion relation (e.g. \citealt{Handbook}). The term $\Delta t_{\rm samp}$ is equal to the actual sampling time for low DMs, and reflects downsampling for searching higher DMs, where the scattering broadening $\tau_{\rm s}$ dominates. We calculate $\tau_{\rm s}$ using the empirical fit of observed scattering versus DM from \cite{Bhat04}. 

%The dispersion smearing over one channel
%\be
%%\Delta t_{\rm DM, 1ch} = \frac{8.3 \mu s DM \Delta\nu_{\rm 1ch, MHz}}{f_{\rm center, GHz}^3},
%\Delta t_{\rm DM} \approx 4.15~{\rm ms}~DM \left(\frac{1}{f_{\rm lo, GHz}^2} - \frac{1}{f_{\rm hi, GHz}^2}\right), \label{eqn_dm}
%\ee
%where DM is the dispersion measure. To calculate the dispersion smearing over a channel width, $\Delta t_{\rm DM, 1ch}$, we use $f_{\rm hi,lo} = f_{\rm center} \pm \Delta\nu_{\rm 1ch}/2$, where $\Delta\nu_{\rm 1ch}$ is the width of one channel. To calculate $\Delta t_{\rm \delta DM}$, the dispersion smearing due to DM error in the worst case, when the pulsar DM is halfway between two trial DMs, we use $f_{\rm hi,lo} = f_{\rm center} \pm \Delta\nu/2$ and replace DM in Equation~\ref{eqn_dm} with the maximum DM error, $\delta$DM, corresponding to half the spacing in the trial DM list. 

\begin{figure}
\centering
\includegraphics[width=\textwidth]{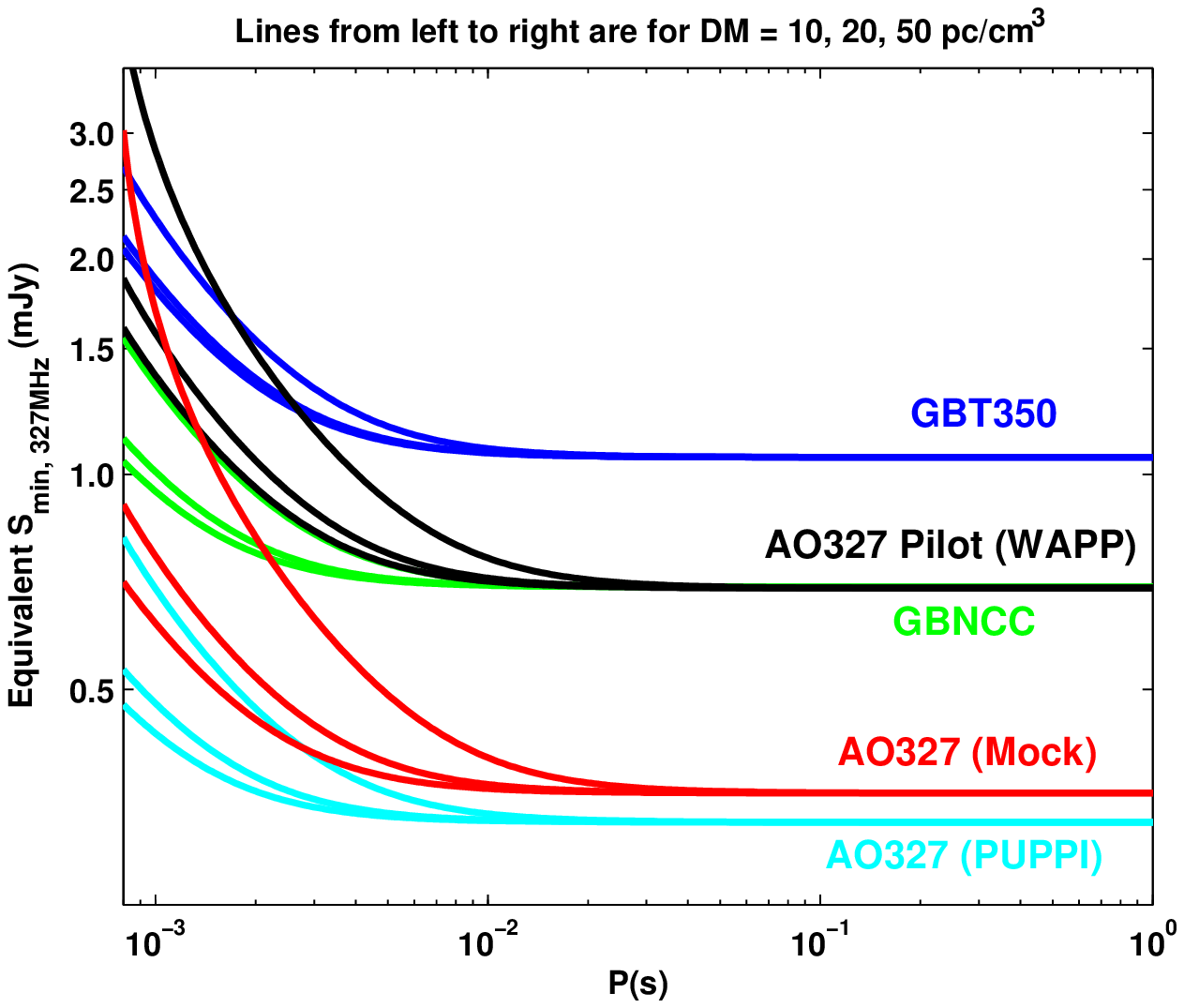}
\caption{Minimum detectable flux density at beam center versus pulse period for the past (WAPP), present (Mock), and future (PUPPI) incarnations of AO327, as well as the GBT350, and GBNCC surveys for three DMs typical of pulsars at high Galactic latitudes. The sensitivity of GBT350 and GBNCC was scaled to 327~MHz using a spectral index $\alpha = -1.7$.\label{fig_sensitivity}}
\end{figure}

%\begin{equation}
%\log  {\tau_{\rm s}(\rm ms)}  \, \simeq \,
%-6.46\,+\,0.154 \log {\rm DM}+\, \,
%1.07 \left (\log {\rm DM} \right)^2
%\,-\,3.86 \log f_{\rm center, GHz}. \label{eqn_taus}
%\end{equation}
In order to compare the boresight sensitivity of several low-frequency, high-latitude surveys, we use the parameters in Table~\ref{tab_surveys}. We scale the GBT350 and GBNCC sensitivity to 327~MHz using a spectral index $\alpha = -1.7$. The current, Mock spectrometer-based version of AO327 is significantly more sensitive than GBT350 or GBNCC to MSPs at low DMs and matches GBT350 sensitivity at DM~$\sim 50$~pc~cm$^{-3}$ (Figure~\ref{fig_sensitivity}). With the new PUPPI backend, AO327 will be more sensitive to MSPs at any DM than either GBT350 or GBNCC. We expect new MSP discoveries to have relatively low DMs, as according to \cite{Bhat04}, for a DM of 100~pc~cm$^{-3}$ and $f_{\rm center} = 327$~MHz, $\tau_{\rm s} = 1$~ms. The detectability of high-DM MSPs in low-frequency surveys is limited by scattering. 

\begin{figure}[t]
\centering
\includegraphics[width=\textwidth,trim= 0 50 0 0, clip=true]{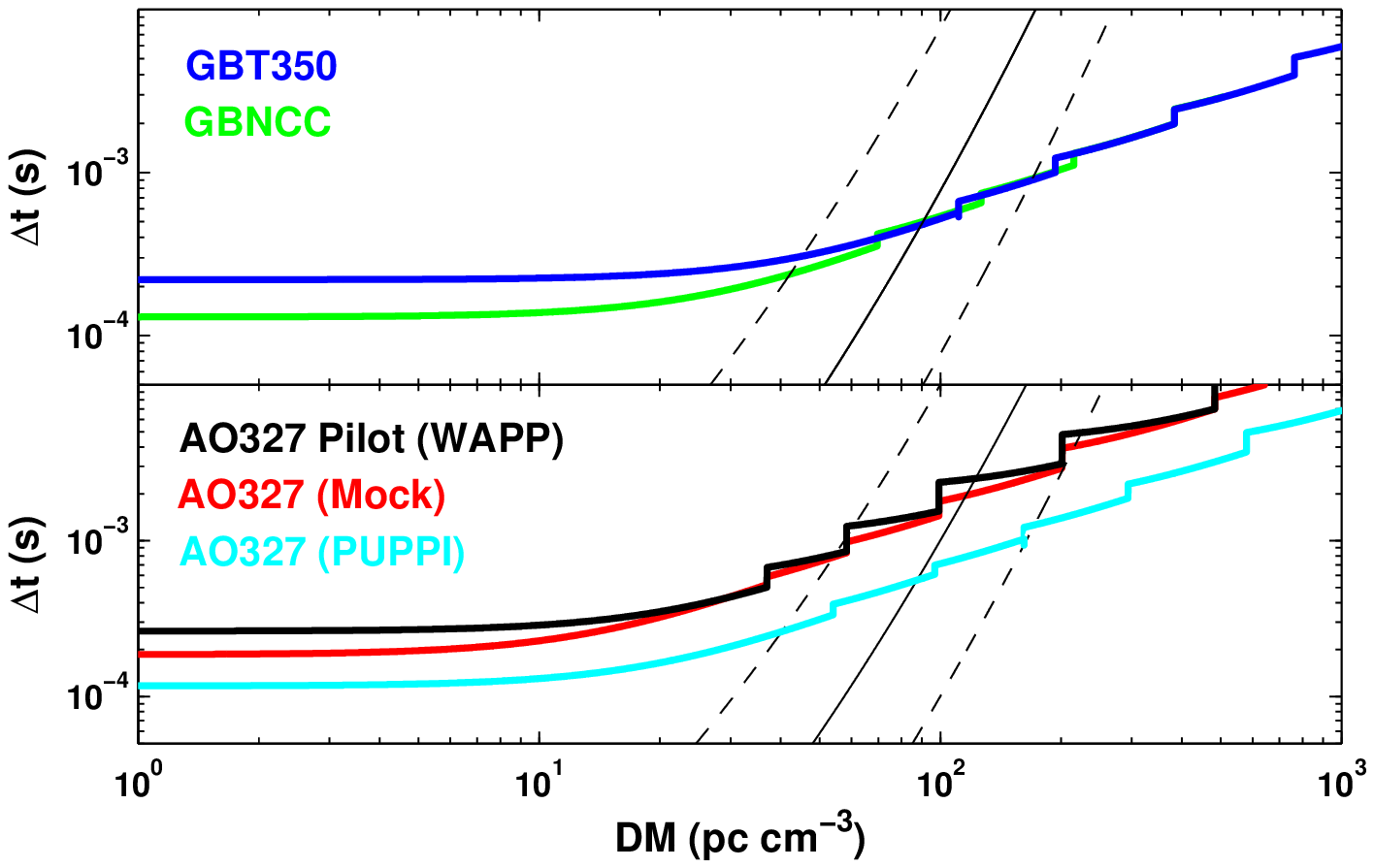}
\caption{Illustration of the cross-over point between a dispersion-limited and scattering-limited detection regime for the surveys we compare in this paper. In both panels, a thin solid black line shows the uncorrectable scattering broadening $\tau_{\rm s}$ as a function of DM according to \cite{Bhat04}. Thin dashed black lines show $\pm$ one order of magnitude from $\tau_{\rm s}$, which is approximately the scatter around this empirical relation. In the top panel, a blue line shows the correctable broadening $\Delta t = \sqrt{\Delta t_{\rm DM, 1ch}^2 + \Delta t_{\rm \delta DM}^2 + \Delta t_{\rm samp} ^2}$ for GBT350, and a green line shows the same quantity for GBNCC. In the bottom panel, $\Delta t$ is denoted by a thick black line for the pilot AO327 survey, and by red and cyan lines for AO327$_{\rm Mock}$ and AO327$_{\rm PUPPI}$, respectively. For trial DMs lower than the DM where $\Delta t({\rm DM})$ and $\tau_{\rm s}({\rm DM})$ cross for a survey, the survey is dispersion-limited. For trial DMs above that point, it is scattering-limited. The plot takes into account downsampling factors and DM spacings, which increase discretely as trial DMs increase. This accounts for the steps in $\Delta t$. AO327$_{\rm PUPPI}$ matches closely the cross-over point and the $\Delta t({\rm DM})$ dependence of GBNCC, the more sensitive of the GBT surveys, while having the advantage of a lower frequency, where pulsars are brighter. \label{fig_dtdm}}
\end{figure}

Another important quantity to consider is the cross-over point between a dispersion-limited and a scattering-limited detection regime as a function of trial DM. In the dispersion-limited regime, we can gain sensitivity to fast pulsars by minimizing the correctable broadening due to the sampling time, downsampling of the data during processing, and dispersion broadening caused by the finite width of a frequency channel and the finite spacing between successive DMs in the trial DM list, $\Delta t = \sqrt{\Delta t_{\rm DM, 1ch}^2 + \Delta t_{\rm \delta DM}^2 + \Delta t_{\rm samp} ^2}$. In the scattering-limited regime, uncorrectable scattering broadening limits the detectability of fast pulsars and improvements in the datataking setup or hardware give diminishing returns. Figure~\ref{fig_dtdm} shows this cross-over point for the surveys in Table~\ref{tab_surveys}. Ideally, the smallest practical DM spacing and no downsampling should be used for search DMs smaller than the cross-over DM. In practice, some sensitivity is sacrificed in favor of shorter data processing times. For search DMs larger than the cross-over DM, downsampling factors and DM spacings are chosen such that $\Delta t << \tau_{\rm s}$.  

\subsection{Search Volume}
\begin{figure}[t]
\centering
\includegraphics[width=0.9\textwidth]{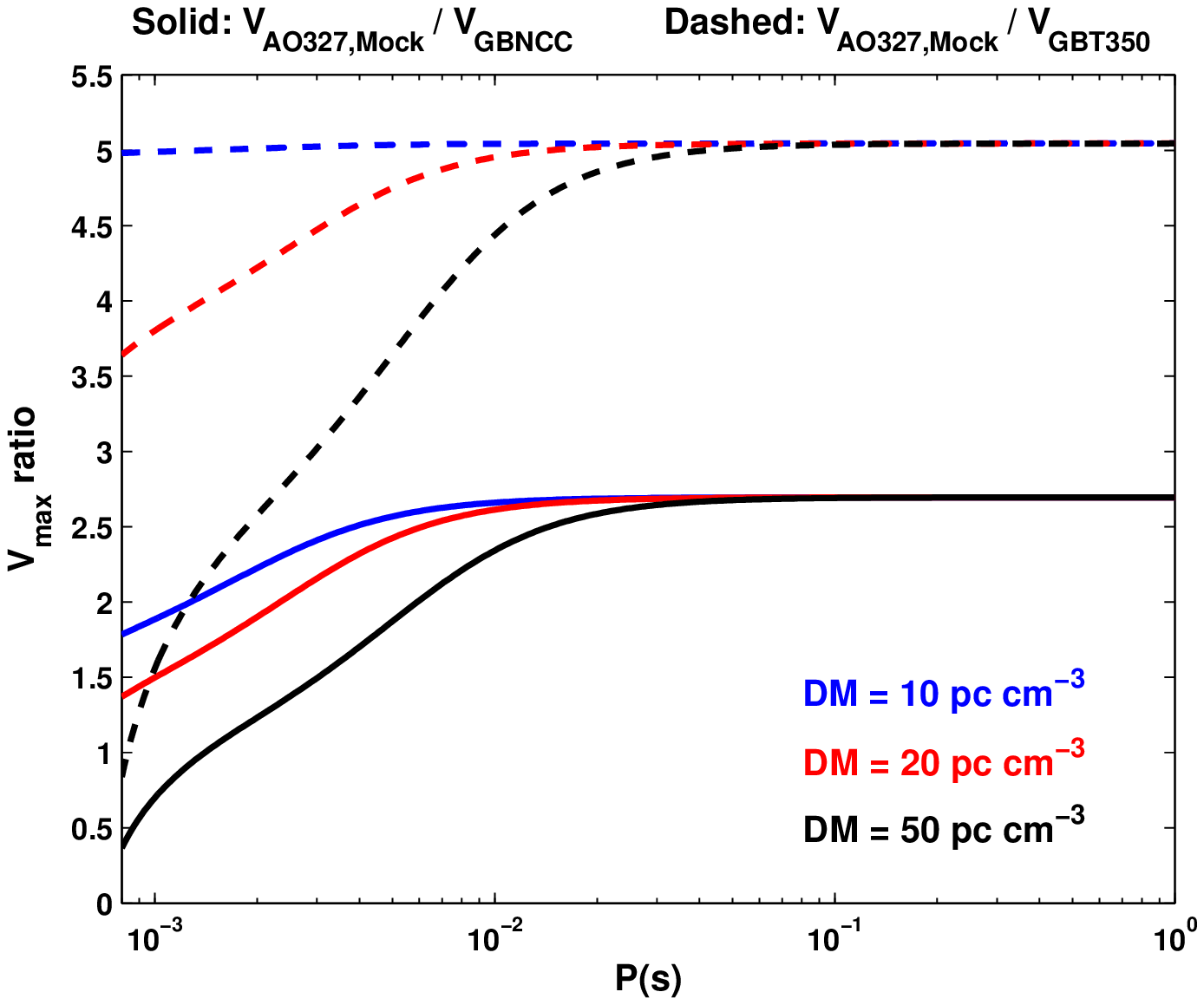}
\caption{Ratios of maximum volume per unit solid angle searched (Equation~\ref{eqn_volume}) for three DMs typical of pulsars at high Galactic latitudes. Solid lines denote $V_{\rm max, AO327-Mock} / V_{\rm max, GBNCC}$ and dashed lines denote $V_{\rm max, AO327-Mock} / V_{\rm max, GBT350}$ for the Mock spectrometer setup currently used for AO327 observations. \label{fig_volume_mock}}
\end{figure}

\begin{figure}[t]
\centering
\includegraphics[width=0.9\textwidth]{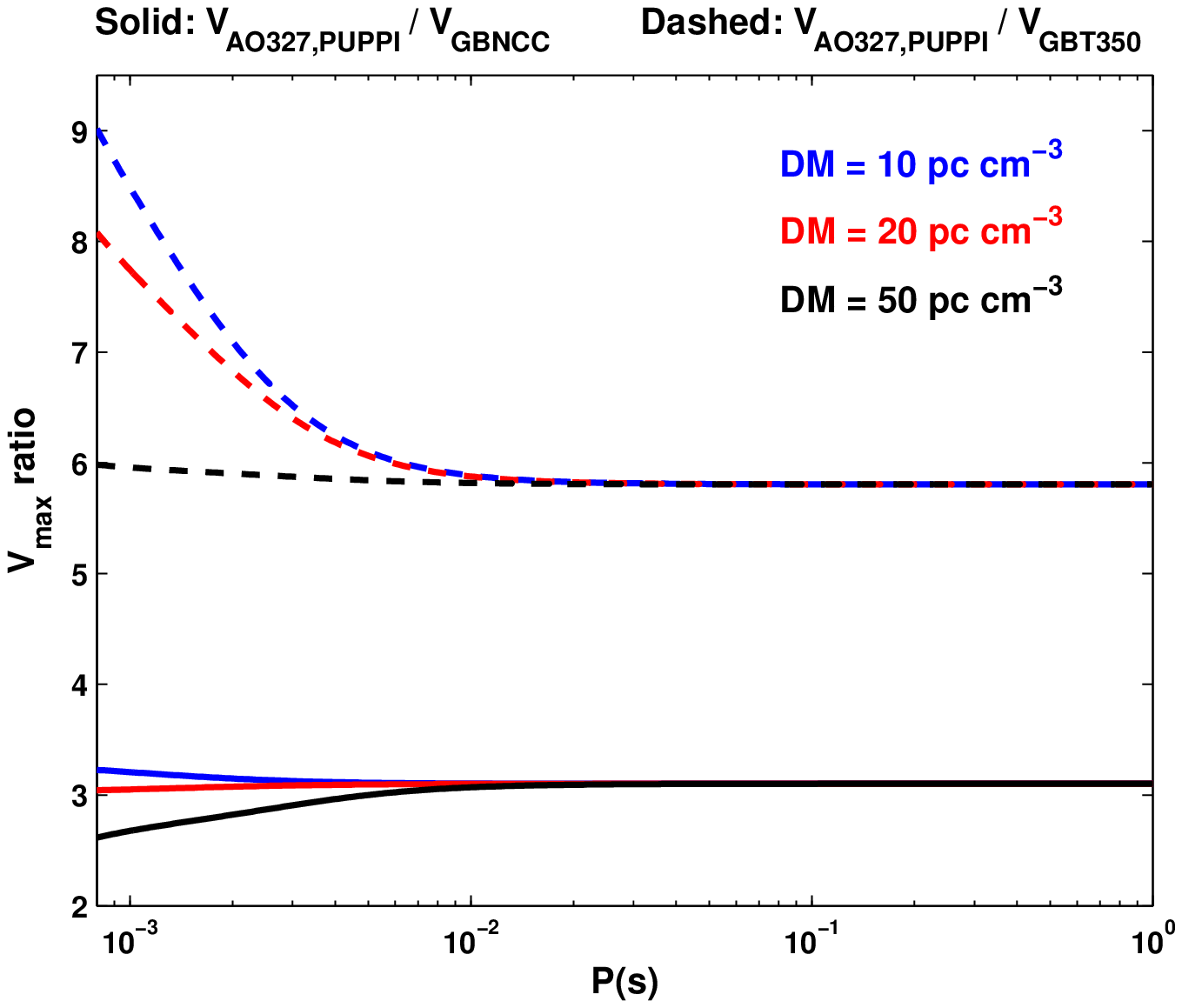}
\caption{Ratios of maximum volume per unit solid angle searched (Equation~\ref{eqn_volume}) for three DMs typical of pulsars at high Galactic latitudes. Solid lines denote $V_{\rm max, AO327-PUPPI} / V_{\rm max, GBNCC}$ and dashed lines denote $V_{\rm max, AO327-PUPPI} / V_{\rm max, GBT350}$ for the new PUPPI backend that AO327 will transition to in the second half of 2013. \label{fig_volume_puppi}}
\end{figure}

\newcommand{\dmax}{d_{\rm max}}
\newcommand{\Vmax}{{V_{\rm max}}}
\newcommand{\Smin}{{S_{\rm min}}}
A key question to consider is how much volume is searched in a survey.
The search volume is a strong function of frequency, direction and
distance and is affected by propagation effects (dispersion and
scattering), search sensitivity $\Smin$ and solid angle $\Omega$.
It is this volume that largely determines the total number of objects
detectable, $N$. Assuming that the survey depth does not
reach the edge of the pulsar population, but a smaller distance
$\dmax$, then $N \propto \Vmax = \frac{1}{3}\Omega d_{\rm max}^3$.
Here, $V_{\rm max}$ is the maximum spatial volume probed. For a given
pulsar at distance $D$ with flux density $S$, $\Vmax \propto D^3
(S/\Smin)^{3/2}$. Comparing AO327 and GBT350, the ratio of survey
volumes is
\begin{equation}
\frac{V_{\rm max, AO327}}{V_{\rm max, GBT350}} = 
		\left [
		\frac
		{
		  S_{\rm 327 MHz}/S_{\rm 350 MHz}
		}
		{
		  S_{\rm min, AO327}/S_{\rm min, GBT350}
		}	
		\right ]^{3/2}.\label{eqn_volume}
\end{equation}
For pulsars with median spectral index $\alpha = -1.7$, the numerator
$S_{\rm 327 MHz}/S_{\rm 350 MHz} = 1.12$. Figure~\ref{fig_volume_mock} shows $V_{\rm max, AO327,Mock}/V_{\rm max, GBT350}$ and $V_{\rm max, AO327,Mock}/V_{\rm max, GBNCC}$ versus pulsar period. For long-period pulsars, whose detectability is not affected by scattering and dispersion, AO327$_{\rm Mock}$ is searching a 4 times larger volume per unit solid angle than GBT350 and a 2.7 times larger volume per unit solid angle than GBNCC. For DMs typical of expected high-latitude MSP discoveries, AO327$_{\rm Mock}$ is searching $\sim 0.7 - 4$ times the volume of GBT350 and $\sim 0.4 - 1.8$ times the volume of GBNCC. The advantages of the GBT surveys in the case of detecting very fast, higher DM pulsars come from their smaller sampling times and narrower channels compared to AO327$_{\rm Mock}$ (Table~\ref{tab_surveys}), as well as the slightly less pronounced scattering and dispersion broadening at 350~MHz compared to 327~MHz. However, that advantage will be drastically reversed in favor of AO327 once we start using the new PUPPI backend (Figure~\ref{fig_volume_puppi}). AO327$_{\rm PUPPI}$ will search a 3.1 times larger volume per unit solid angle than GBNCC and a 4.7 times larger volume than GBT350 for slow pulsars. For MSPs, AO327$_{\rm PUPPI}$ will search $\sim 2.6 - 3.3$ times the volume of GBNCC and $\sim 4.8 - 7.2$ times the volume of GBT350. 

\subsection{Data Processing}

Data are recorded by the Mock spectrometer as 16-bit short integers in the PSRFITS data format. We have determined that 4 bits per data point provide enough dynamic range so that sensitivity is unaffected, and we routinely rescale and rewrite data files in a 4-bit PSRFITS format for shipping to processing sites and for long-term storage in order to minimize disk needs. Data are processed at the University of Texas at Brownsville (UTB), West Virginia University (WVU), and the Arecibo Observatory (AO) using a pipeline based on the PRESTO pulsar search code\footnote{\tt http://www.cv.nrao.edu/~sransom/presto}. The AO beam width at 327~MHz is 15\amin, giving a maximum transit time of one minute. Data are processed in 60-second spans, overlapping successive spans by 30 seconds. Data are dedispersed with 6358 trial DMs in the range $0 - 1000$~pc~cm$^{-3}$ and searched for periodic and transient signals. We defer a more detailed and technical discussion of the search pipeline to a future paper. 

Periodic and single-pulse candidates are stored in databases at UTB and WVU. A candidate ranking method based on the PEACE algorithm \citep{Lee13} assigns a grade to each periodic candidate corresponding to the candidate's likelihood of being a real pulsar. The grade determines in what order candidates are examined by participants from AO, UTB, WVU, and the University of Wisconsin at Milwaukee (UWM) via web portals hosted at UTB and WVU. There are currently more than 50 AO327 participants with various levels of involvement in the survey and expertise, ranging from high school students to university professors.

%--List of known pulsars detected along with offsets from beam center and profile SNR? (Include S400 from ATNF along with estimated flux at 327MHz based on actual or average spectral index). 

%--Plot of S vs. P of known and new detections along with Smin vs. P for survey? Or include dots for pulsars in sensitivity curves plot? (There all curves are rescaled to 327MHz using the average spectral index). 

\section{First Discoveries}\label{sec_new}
So far the AO327 search pipeline has detected 44 known pulsars with periods in the range $0.003 - 2.21$~s 
%{\bf [XXX update count and range before submitting]} 
and discovered 24 pulsars (Table~\ref{tab_new}), including three MSPs, three objects with periods typical of young as well as mildly recycled pulsars, a nuller, and a rotating radio transient (RRAT). Five of the pulsars in Table~\ref{tab_new} are included in an Arecibo timing campaign that began in February 2013. We expect to begin timing the rest of the new discoveries in mid to late 2013. Four of the new pulsars are within 10\degrees\ of the Galactic plane. Two of these pulsars are close to the portion of the inner Galaxy visible from Arecibo, and one, PSR~J1821+01, has $P \sim 34$~ms. PALFA searches the region within $5\degrees$ of the Galactic plane in the portion of the Galaxy accessible to the Arecibo telescope: $32\degrees < l < 77\degrees$ and to a lesser extent $168\degrees < l < 214\degrees$. The fact that a low-frequency survey like AO327 is finding fast pulsars close to the inner Galactic plane underscores the necessity for mid-latitude targeted pulsar searches in this part of the sky. Ideally future surveys of the $5\degrees \lesssim |b| \lesssim 15\degrees$ region will use frequencies higher than 327~MHz in order to minimize scattering and make it possible to probe deeper parallel to the Galactic plane, but lower than the PALFA frequency of 1.4~GHz, where pulsars are significantly weaker and the number of pointings required would be prohibitively large. 

\begin{figure}[t]
\centering
\includegraphics[width=0.9\textwidth]{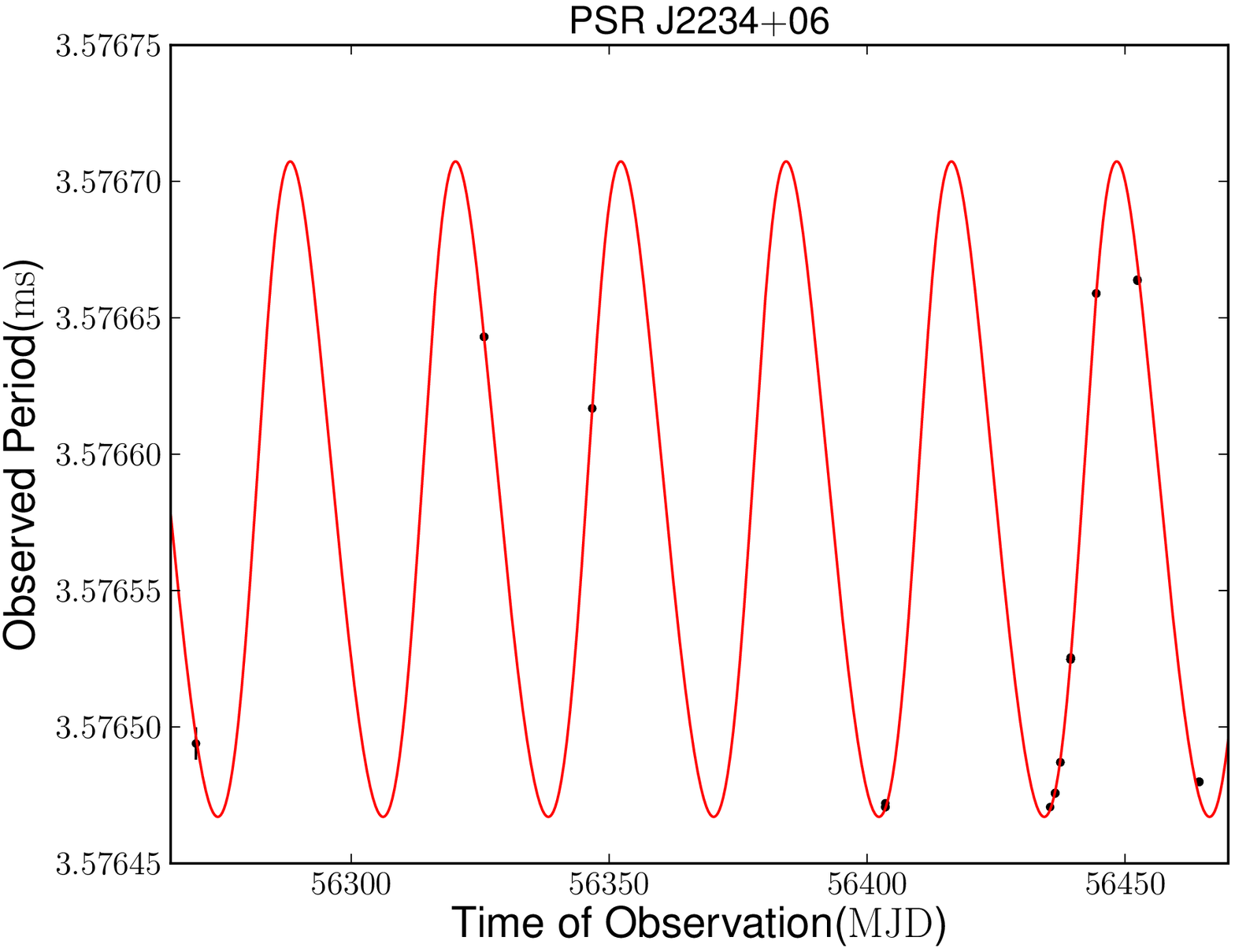}
\caption{Barycentric period measurements of PSR~J2234+06 vs epoch. Black dots denote observations and the red curve shows the expected change in observed period due to eccentric orbital motion for the best-fit orbit. In this case, the orbital period is 32 days, and $e = 0.13$. \label{fig_2234orbit}}
\end{figure}

\begin{figure}[t]
\centering
\includegraphics[width=0.9\textwidth]{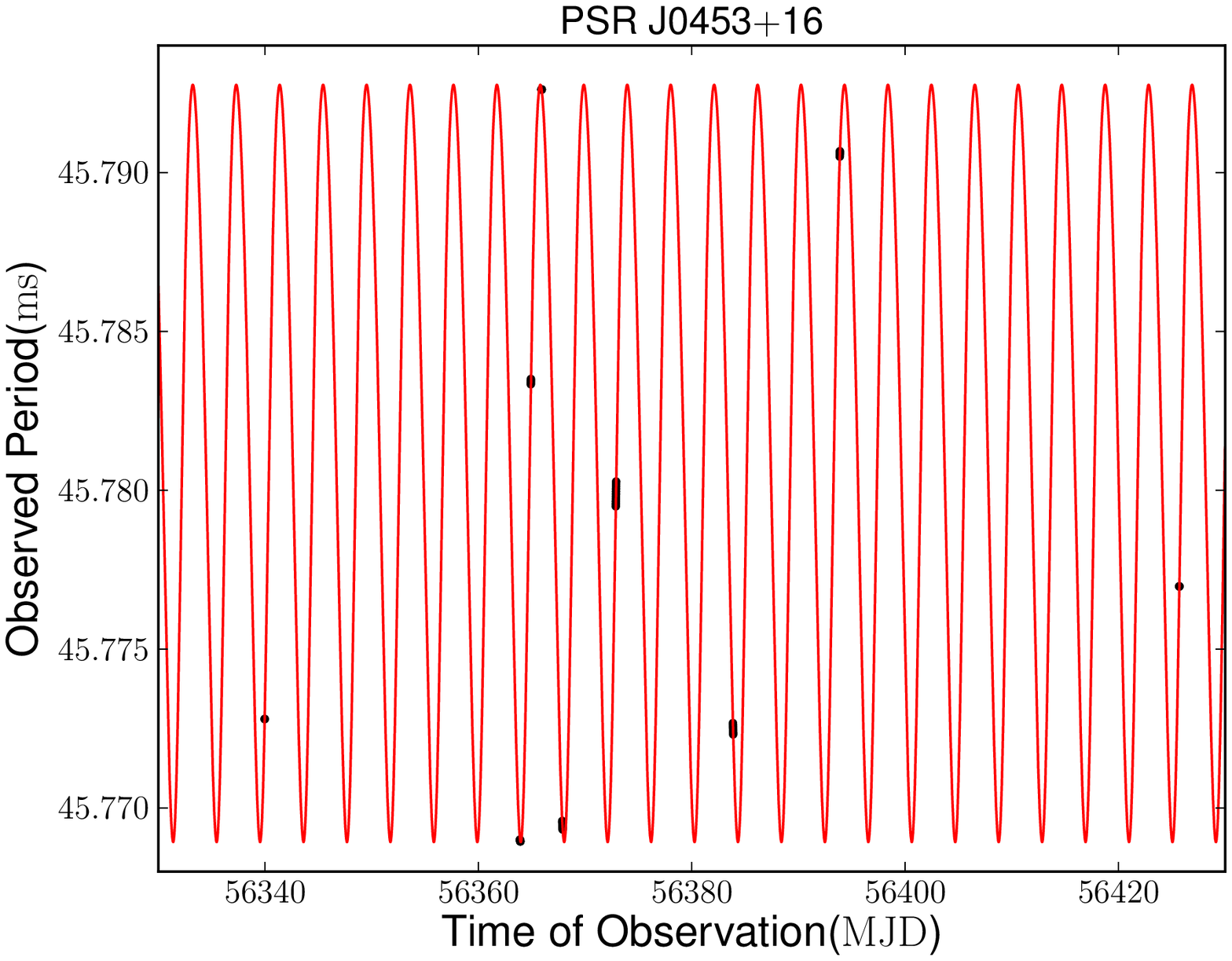}
\caption{Barycentric period measurements of PSR~J0453+16 vs epoch. Black dots denote observations and the red curve shows the expected change in observed period due to eccentric orbital motion for the best-fit orbit. In this case, the orbital period is 4 days, and $e = 0.11$. \label{fig_0453orbit}}
\end{figure}

\subsection{PSR~J2234+06}

PSR~J2234+06 is an MSP with $P = 3.58$~ms and a 32-day orbit (Fig.~\ref{fig_2234orbit}). Assuming a pulsar mass of 1.4~\msun, the companion star has a minimum and median mass of about 0.19 and 0.23~\msun, respectively. The orbital eccentricity of this system, $e = 0.13$, is much higher than observed for most other Galactic millisecond pulsars, which have orbits with eccentricities on the order of $10^{-7}$ to $10^{-3}$. This trend is generally understood to be an inevitable consequence of orbital circularization during the accretion episode that leads to the recycling of the pulsar (e.g., \citealt{PK94}). The spin and orbital parameters of PSR~J2234+0611 make it a more compact, less massive cousin to PSR~J1903+0327, an MSP with $P = 2.15$~ms, $P_{\rm b} = 95$ days, $e = 0.44$, and a main sequence companion with a mass of $\sim 1~\msun$ \citep{Champion08}. Such systems cannot be formed within current understanding of stellar evolution and are believed to originate either from an exchange interaction in a region with a high stellar density or from a hierarchical triple that becomes dynamically unstable due to mass transfer to the pulsar from a third system component that is eventually ejected. \cite{Freire11} reach the latter conclusion about PSR~J1903+0327 based on the motion of the system in the Galaxy and \cite{PZ11} independently reach the same conclusion based on numerical simulations of the evolution of the system.

Follow-up observations of PSR~J2234+06 at 1.4~GHz revealed that the pulsar is very bright and can be easily timed at this frequency. Based on the folded pulse profiles at 1.4~GHz and 327~MHz, we estimate a peak flux density of 15~mJy at 1.4~GHz and 4.4~mJy at 327~MHz. The corresponding period-averaged flux densities are 0.7 and 0.3~mJy, respectively. This suggests that the pulsar either scintillates strongly on a time scale of minutes or has an anomalous spectrum rising with frequency. Most pulsars exhibit power-law spectra, $S \propto \nu^\alpha$, with an average spectral index $\alpha = -1.7$, and MSPs spectra on average decline even more steeply with increasing frequency. PSR~J2234+06 has the lowest DM of all AO327 discoveries so far, which makes it especially prone to scintillation. One 1.4~GHz timing observation (25 March 2013) did not produce a detection. We rule out instrumental problems as other pulsars observed with the same setup before and after PSR~J2234+06 were successfully detected in the same session. The pulsar was again detected easily at 1.4~GHz in subsequent timing observations. Apart from scintillation, another possible explanation of the one non-detection is that PSR~J2234+06 is in an eclipsing binary system. 

Time-of-arrival (TOA) measurements extracted from our 1.4~GHz observations have uncertainties on the order of only 0.4~$\mu$s within 2.5 minutes. Since $\sigma_{\rm TOA} \propto 1/\sqrt{T_{\rm obs}}$, we expect TOAs of this pulsar to have uncertainties on the order of $0.11~\mu$s within half an hour, comparable to the typical precisions achieved within a half-hour observation for pulsars included in the PTAs now being used to search for gravitational waves (\citealt{Demorest13}, \citealt{Hobbs10}). 

The brightness of PSR~J2234+06 at 1.4~GHz and the orbital eccentricity imply that, as in the case of PSR~J1903+0327, we will certainly be able to measure the rate of advance of periastron ($\dot{\omega}$). If the companion is compact, $\dot{\omega}$ is solely due to the effects of general relativity and can be used to estimate the total mass of the system \citep{WT81}. If the orbit is highly inclined with respect to the plane of the sky, the excellent timing precision of PSR~J2234+06 means we will likely be able to measure the Shapiro delay, and therefore obtain a mass ratio and determine the masses of both components. 

\subsection{PSR~J0453+16}

PSR~J0453+16 has $P = 45.8$~ms and DM~$= 30$~pc~cm$^{-3}$. The discovery observation was made on 30 May 2010 and the pulsar was confirmed on 21 August 2012. The effective barycentric period derivative between these two observations is $\dot{P}_{\rm eff} = -1.47(7) \times 10^{-13}$~s/s. Several subsequent observations show sign changes in $\dot{P}_{\rm eff}$. Since rotation-powered pulsars typically spin down as energy is transferred to the emission mechanism, we conclude that the change in period observed for this pulsar is affected by orbital motion in a binary system. A fit for the orbit (Fig.~\ref{fig_0453orbit}) indicates an orbital period of 4 days and an eccentricity of 0.11. The projected semi-major axis is 14.5~lt-s. Using these parameters and assuming a pulsar mass of 1.4~\msun, we get a minimum and mean companion masses of 1.0~\msun\ and 1.3~\msun, respectively. Given its high eccentricity, mean companion mass of about 1.3~\msun, and distance from the galactic plane ($\sim 0.32$~kpc), PSR~J0453+16 may be in a double neutron star system. This pulsar is also easily detectable at 1.4~GHz and along with the orbital eccentricity this implies that, as the case of PSR~J2234+06, we will be able to measure $\dot{\omega}$ and estimate the total mass of the system.

\subsection{PSR~J0509+08}

PSR~J0509+08 has $P = 4.06$~ms and DM~$= 38$~pc~cm$^{-3}$. The pulsar was discovered in an observation from 2 May 2010 and confirmed on 19 April 2013. We use the barycentric periods and respective uncertainties reported by Presto for these data sets to calculate the effective barycentric period derivative between the two observations, $\dot{P}_{\rm eff} = 2.83(3) \times 10^{-15}$~s/s. MSPs typically have period derivatives on the order of $10^{-20} - 10^{-19}$~s/s. Of 151 pulsars with $P < 0.015$~s and measured $\dot{P}$ in the ATNF catalog, only two have $\dot{P} \sim 10^{-18}$~s/s and none have a larger $\dot{P}$. We therefore conclude that the observed period change for PSR~J0509+08 is due to its orbital motion in a binary system. We estimate the flux density of this pulsar to be 5~mJy at 327~MHz and 0.8~mJy at 1.4~GHz, corresponding to a spectral index of $-1.3$. A TOA measurement extracted from a 20-minute follow-up observation at 327~MHz has uncertainty of only 0.2~$\mu$s, indicating that PSR~J0509+08, like PSR~J2234+06, is an excellent candidate for inclusion in PTAs. A TOA extracted from a 4-minute observation at 1.4~GHz has uncertainty of 3.7~$\mu$s, corresponding to a projected TOA uncertainty of 1.3~$\mu$s within half an hour. We expect the timing precision at 1.4~GHz, a frequency typically used for PTA timing, to improve after we obtain a more precise position from regular timing observations and can observe the pulsar at boresight.

\subsection{PSR~J0824+00}

PSR~J0824+00 has $P = 9.86$~ms and DM~$= 35$~pc~cm$^{-3}$. For this pulsar, $\dot{P}_{\rm eff} = 2.4(1) \times 10^{-14}$~s/s between the discovery observation from 2 September 2012 and the confirmation from 16 April 2013. This suggests orbital motion for the same reasons as for PSR~J0509+08.

\subsection{PSR~J2204+27}

PSR~J2204+27 has $P = 84.7$~ms and DM~$= 35$~pc~cm$^{-3}$. $\dot{P}_{\rm eff}$ between the discovery observation on 22 May 2010 and the confirmation on 31 August 2011 is $1.8(3) \times 10^{-13}$~s/s. Five observations at 327~MHz within 24 days in September 2011 do not show a change in the observed period to within the precision reported by Presto. However, the effective period derivative between September 2011 and the start of our timing campaign in February 2013 is $-4.7(9) \times 10^{-14}$~s/s, again pointing to binary motion as the explanation. Since no $\dot{P}_{\rm eff}$ was observed on a timescale of days, this suggests that the orbital period is on the order of months or longer. 
%{\bf [XXX Can we confidently say that? If so, what other implications, if any, are there? E.g. with a P=85ms, the pulsar must be either youngor mildly recycled; if the latter, in order for the pulsar to have been able to accrete matter, the system must be either eccentric (to bring it close enough to the companion for accretion), or undergone some interaction in the past?]}

\subsection{PSR~J1821+01}

PSR~J1821+01 has $P = 33.8$~ms and DM~$= 52$~pc~cm$^{-3}$. The calculated uncertainty in $\dot{P}_{\rm eff}$ is $1.1 \times 10^{-13}$~s/s between the 24 September 2012 discovery and the 3 November 2012 confirmation, and $1.1 \times 10^{-15}$~s/s between the confirmation and our first timing observation from 18 April 2013. We did not detect any $\dot{P}_{\rm eff}$ significantly larger in magnitude than these uncertainties. The period of PSR~J1821+01 is consistent with either a young or a mildly recycled pulsar. Young pulsars have $\dot{P} \sim 10^{-14} - 10^{-12}$~s/s, and mildly recycled pulsars have $\dot{P} \sim 10^{-17} - 10^{-19}$~s/s. Our lack of detection of a $\dot{P}_{\rm eff}$ even between observations separated by 166 days suggests that PSR~J1821+01 is an isolated mildly recycled pulsar. PSR~J1821+01 has $b \approx 7\degrees$, and the catalog of Galactic supernova remnants \citep{Green09} has no nearby entries. The X-ray source 1RXS~J182217.2+014227 is at the edge of the discovery beam, $\sim 7.5$\amin\ from beam center. We ruled out association with the pulsar by pointing at the X-ray source immediately after the successful confirmation observation and observing it with the same 327~MHz telescope setup and 10-minute integration time. This pointing produced no detection.

\subsection{PSR~J0229+20}

\begin{figure}
\centering
\includegraphics[angle=-90]{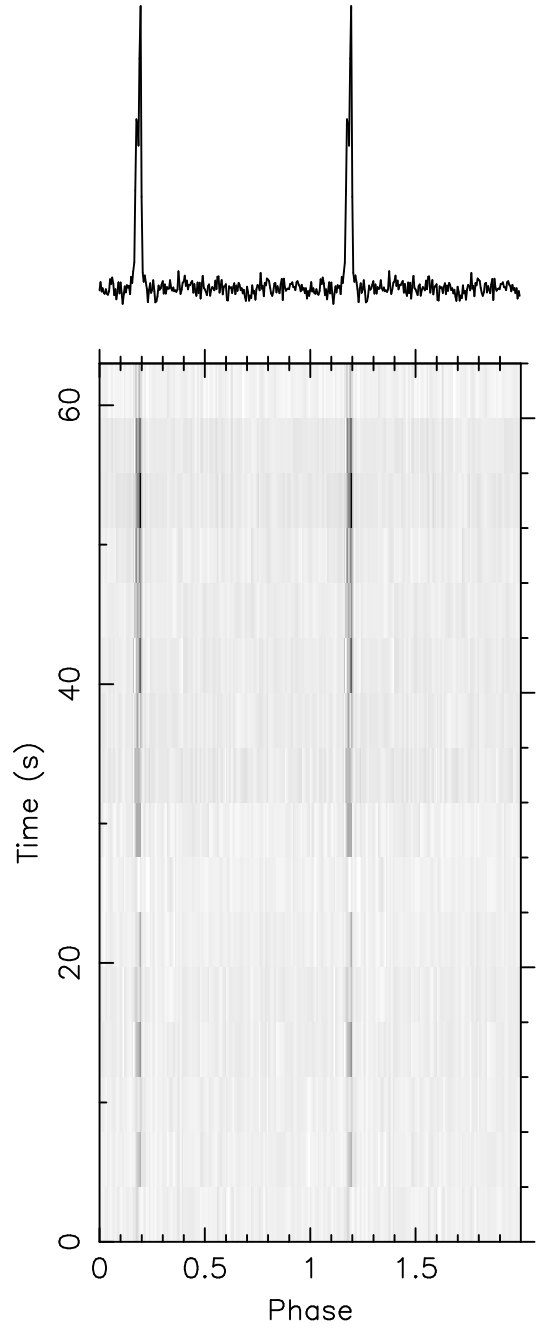}
\caption{Folded pulse profile (top) and pulse phase vs. time (bottom) for the discovery observation of PSR~J0229+20. A gap in emission is evident at $t \sim 25$~s and may be due to nulling or scintillation.\label{fig_0229prepfold}}
\end{figure}

PSR~J0229+20 has $P = 806.9$~ms and DM~$= 27$~pc~cm$^{-3}$, and was discovered via a periodicity search. Figure~\ref{fig_0229prepfold} shows the pulse profile averaged over the $\sim 60$~s of the discovery observation, as well as a greyscale panel of subintegration vs. pulse phase. The length of a subintegration for the purpose of plotting is partially determined by the period of a pulsar candidate, and in this case each row in the greyscale panel corresponds to an average of $\sim 4$~s of data. Darker regions indicate stronger emission. The observed intensity variations between subintegrations may be affected by intrinsic variability and/or scintillation. The emission in the first subintegration ($t = 0 - 4$~s) of the data span shown is weak as the pulsar enters the half-power contour of the telescope beam. The subintegration at $t \approx 24 - 28$~s shows little to no emission compared to other subintegrations. During the last subintegration ($t \gtrsim 60$~s) the pulsar begins to exit the telescope beam and emission intensity decreases. The beam of the 327~MHz Arecibo receiver is elliptical ($\approx 15\amin \times 14\amin$). In addition, the parallactic angle varies between drift tracks done at different azimuth and zenith angles. These two properties account for the fact that even for steady emitters, the observed intensity vs. time generally does not trace out an ideal Gaussian beam cross-section. 

\begin{figure}
\centering
\includegraphics[width=\textwidth,trim= 0 10 0 65, clip=true]{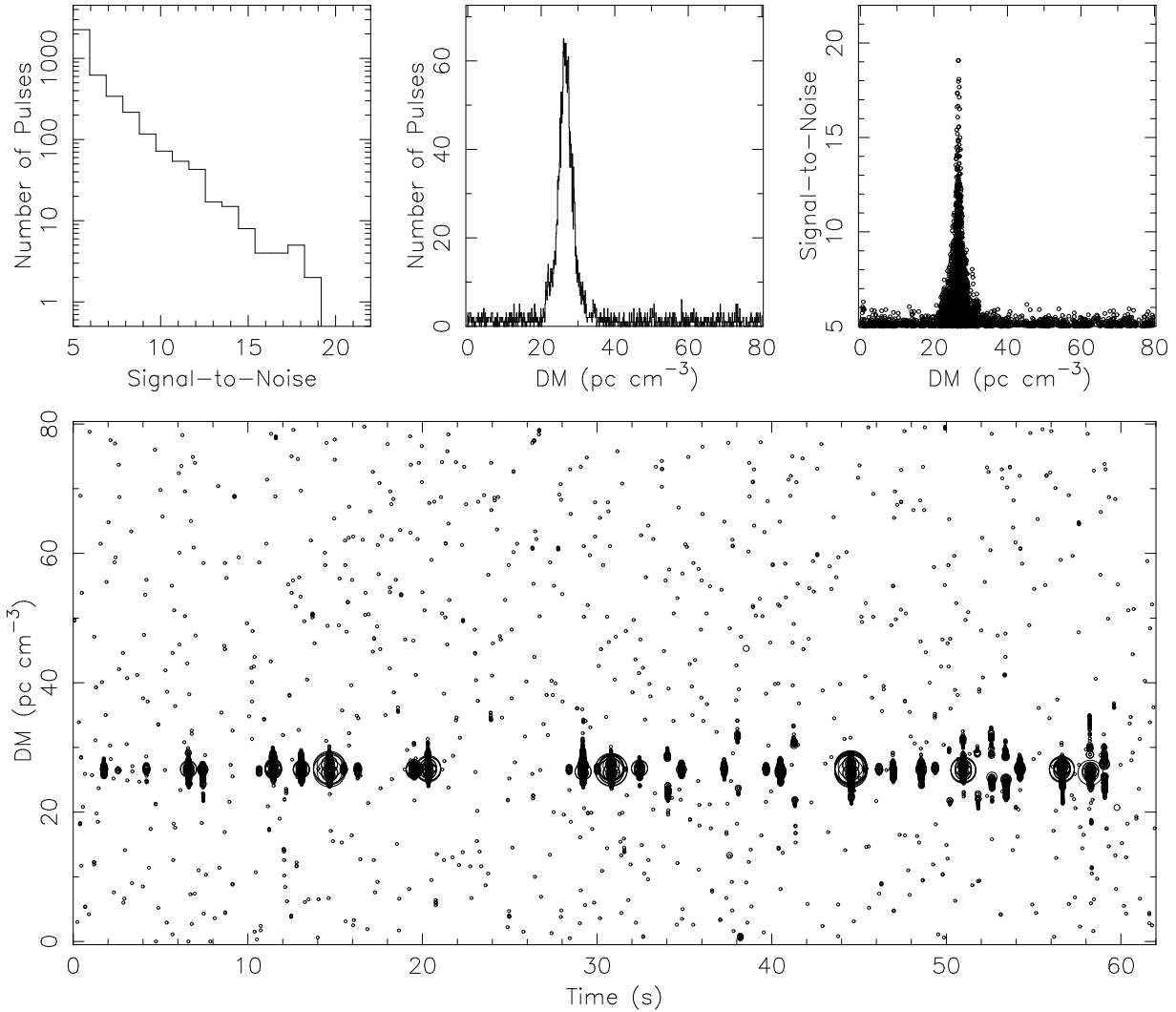}
\caption{Single pulse search output for the discovery observation of PSR~J0229+20. The top panels show the number of pulses vs. SNR and DM, and SNR vs. DM for events with SNR~$> 5$. At bottom the same events are plotted vs DM and time. Brighter events are represented by larger circles. Events are weaker at the start of the data span shown as the pulsar enters the telescope beam. In contrast, the gap at $t \sim 22 - 28$~s has no single pulse events and is flanked by bright pulses before and after. This may be due to nulling or scintillation.\label{fig_0229sp}}
\end{figure}

Figure~\ref{fig_0229sp} shows a plot of the same data span produced by the single pulse branch of our search pipeline. The top three panels show statistics for bursts with ${\rm SNR} > 5$. The prominent peaks in the number of pulses and SNR vs. trial DM plots are indicative of a confident detection of many bright dispersed pulses. The observed single pulse SNR and total pulse count are highest at the trial DM closest to the actual pulsar DM. In the bottom panel, pulse intensity gradually increases at the start of the observation as the pulsar enters the beam. However, there are no obvious non-background events at $t \approx 22 - 28$~s, corresponding to the gap in periodic emission from Figure~\ref{fig_0229prepfold}. Nulling pulsars typically shut off abruptly, and just as abruptly start emitting again. This is consistent with bright single pulses bracketing the absence of detections at $t \approx 22 - 28$~s.

We calculate the diffractive scintillation time scale $\Delta t_{\rm DISS}$ for PSR~J0229+20 based on equations in Section~4 of \cite{Handbook}, the average 2D speed for non-recycled pulsars of $\sim 246$~km/s \citep{Hobbs05}, and a distance estimate of 1.1~kpc based on the pulsar DM and position, and the NE2001 model of ionized gas distribution in the Galaxy. We obtain $\Delta t_{\rm DISS} = 54$~s. However, the empirical relation from \cite{Bhat04} used to calculate the scattering broadening time has a scatter of an order of magnitude, which makes $\Delta t_{\rm DISS}$ values of $17 - 173$~s plausible, and we cannot completely rule out scintillation as the cause of the observed brightness variability of PSR~J0229+20. 

\subsection{PSR~J1010+15}

A different type of pulse intensity modulation is observed in the case of rotating radio transients, which emit pulses at a rate of a few per minute to a few per several hours. After enough pulses are observed, it is usually possible to determine a rotation period before (or without) any timing follow-up observations. Known RRATs appear to have longer periods on average than steadily emitting, non-recycled pulsars (\citealt{McLaughlin06}, \citealt{Deneva09}). However, when the detected number of pulses is small, the large intervals between successive pulses may cause the calculated period to be an integer multiple of the actual period.

\begin{figure}[t]
\centering
\includegraphics[width=\textwidth, trim= 0 10 0 65, clip=true]{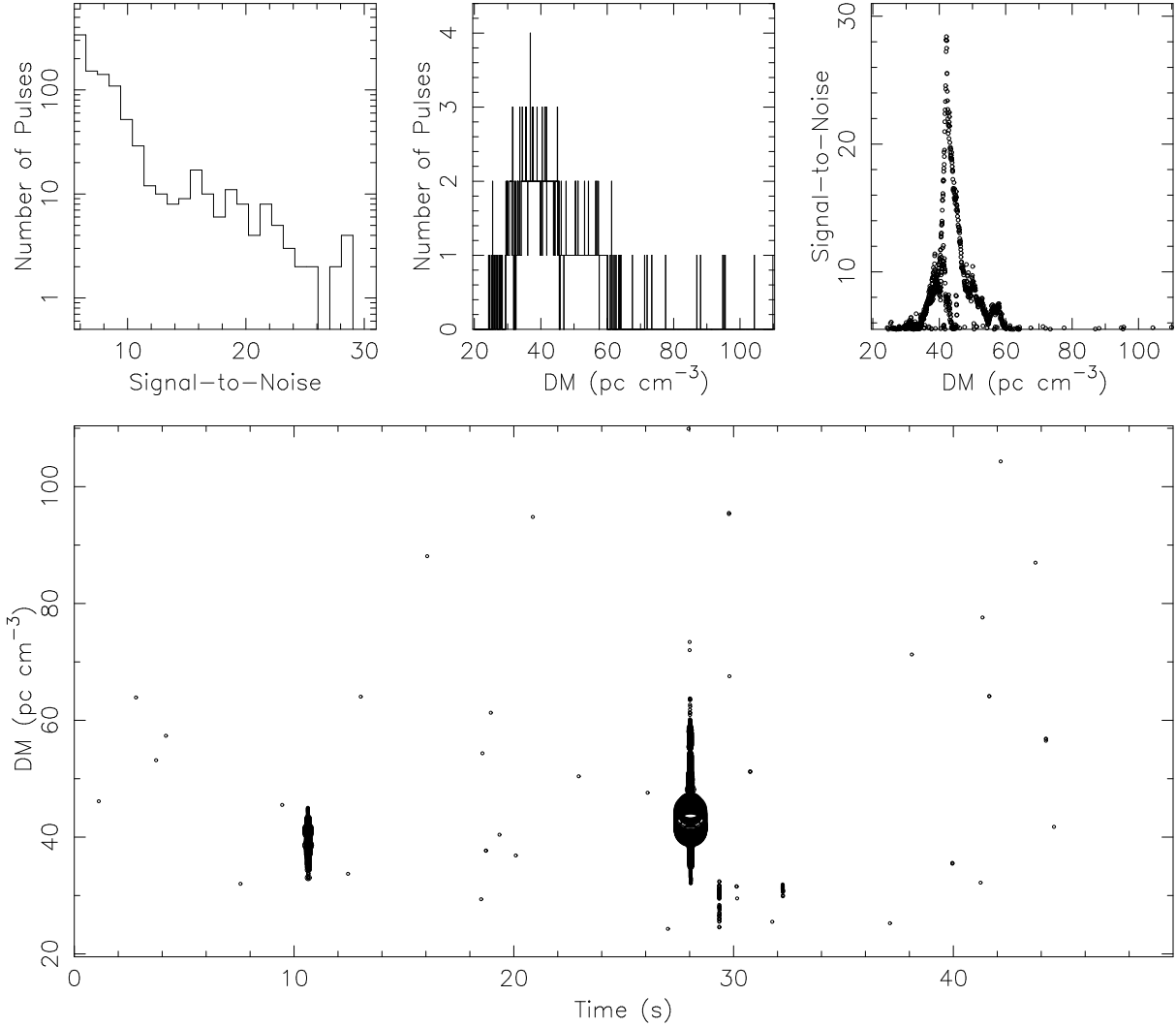}
\caption{Discovery observation of PSR~J1010+15. The top panels show the number of pulses vs. SNR and DM, and SNR vs. DM for events with SNR~$> 5$. At bottom the same events are plotted vs DM and time. Brighter pulses are represented by larger circles. The SNR peaks at DM~$\sim 42$.\label{fig_rrat}}
\end{figure}

As RRATs are a relatively new pulsar class, large surveys offer an opportunity to better understand their population statistics and the causes of their intermittency. In addition, since surveys covering high latitudes like AO327 look away from most of the ionized gas in our Galaxy, they are sensitive to extragalactic radio transients from potentially exotic sources such as merging and annihilating black holes. AO327 has discovered one RRAT to date, PSR~J1010+15. This object has DM~$= 42$~pc~cm$^{-3}$. Figure~\ref{fig_rrat} shows the discovery observation, with two bright dispersed pulses detected $\sim 18$~s apart. There is no corresponding detection from the periodicity search branch of our pipeline.

\begin{table}[t]
\begin{center}
\caption{New pulsars discovered by the AO327 drift survey. All pulsars except PSR~J1010+15 were discovered via periodicity search. The column labeled ``SP'' shows whether an object was also detected in a single-pulse search.\label{tab_new}}
\begin{tabular}{lcccccc}
\hline
Name & RA  & DEC  & $P$ & DM  & SP & Notes \\
& (hh:mm:ss)$^a$ & (dd:mm) & (ms) & (pc~cm$^{-3}$) & & \\
\hline
J0158+21 & 01:58:29 & 21:08 & 505.3 & 20 & & \\
J0229+20 & 02:29:03 & 20:58 & 806.9 & 27 & Y & Nuller \\
J0241+16$^b$ & 02:41:46 & 16:04 & 1545.4 & 16 & Y & \\
J0244+14 & 02:44:51 & 14:27 & 2128.1 & 31 & Y & \\
J0453+16 & 04:53:38(7) & 16:01(1.7) & 45.8 & 30 & & Binary \\
J0457+23 & 04:57:06(7) & 23:34(1.7) & 504.9 & 59 & Y & \\
J0509+08 & 05:09:24(7) & 08:57(1.7) & 4.1 & 38 & & Binary \\
J0608+00 & 06:08:49 & 00:39 & 1076.2 & 48 & Y & \\
J0628+06 & 06:28:04 & 06:38 & 346.5 & 87 & & \\
J0806+08 & 08:06:05 & 08:17 & 2063.1 & 46 & Y & \\
J0824+00 & 08:24:16 & 00:27 & 9.9 & 35 & & Binary\\
J0848+16$^b$ & 08:48:53 & 16:43 & 452.4 & 38 & Y & \\
J0928+06 & 09:28:44 & 06:14 & 2060.4 & 50 & Y & \\
J1010+15$^b$ & 10:10:00 & 15:51 & - & 42 & Y & RRAT \\
J1726$-$00 & 17:26:23 & $-$00:15\phn & 1308.6 & 57 & & \\
J1802+03 & 18:02:44 & 03:38 & 664.3 & 77 & & \\
J1807+04 & 18:07:25 & 04:05 & 798.9 & 53 & Y & \\
J1821+01 & 18:21:48(7) & 01:52(1.7) & 33.8 & 52 & & \\
J1937$-$00 & 19:37:09 & $-$00:17\phn & 240.1 & 68.6 & & \\ 
J1945+07 & 19:45:55 & 07:17 & 1073.9 & 62 & Y & \\
J2204+27 & 22:04:40(7) & 27:02(1.7) & 84.7 & 35 & & Binary\\
J2234+06 & 22:34:22(7) & 06:11(1.7) & 3.6 & 11 & & Binary\\
J2329+16 & 23:29:50 & 16:57 & 632.1 & 31 & Y & \\
J2340+08 & 23:40:45(7) & 08:33(1.7) & 303.3 & 24 & Y & \\
\hline
\end{tabular}
\end{center}
$^a$ RA and DEC are given in the J2000 coordinate system. The uncertainties in both coordinates are 7.5\amin, the 327~MHz beam radius, unless otherwise indicated. The positions with smaller uncertainties were obtained from follow-up observations at 1.4~GHz. \\
$^b$ Found in AO327 Pilot data from 2003. 
\end{table}

\section{Future Directions}\label{sec_future}

We have described the current status and initial set of pulsar discoveries from the AO327 Arecibo drift survey. Phase I of the survey includes sky in the declination range $-1\degrees$ to $28\degrees$, excluding the region with $|b| < 5\degrees$, which is searched by the PALFA consortium at 1.4~GHz, a frequency better suited to the high-scattering and dispersion environment of the Galactic plane. To date we have completed observations of 52\% of the Phase I region, processed 55\% of the data, and discovered 24 new pulsars.
%{\bf [XXX update number]}. 
We expect that most of the remaining Phase I region will be searched with the new PUPPI backend, which provides finer time and frequency resolution, wider bandwidth, and therefore better sensitivity, especially to MSPs. Based on the current time AO327 time allocation of $\sim 400$~hours per year, we expect to complete Phase I of the survey by 2017. Phase II of AO327 will target sky in the declination range $28\degrees - 38\degrees$ and we estimate that will take two additional years to complete. 

Four of the 24 AO327 discoveries have $|b| < 10\degrees$, and one of these, PSR~J1821+01, has $P = 0.034$~s. This pulsar is outside the PALFA search region within 5\degrees of the Galactic plane. The fact that AO327 is finding fast pulsars at relatively low Galactic latitudes suggests that a non-drift search probing deeper parallel to the Galactic plane at $5\degrees \lesssim |b| \lesssim 10\degrees$ will likely be very fruitful and contribute to a more complete picture of the Galactic pulsar population. 

When AO327 discovers a new pulsar, we routinely check for nearby sources from existing optical and high-energy catalogs, including the Second Fermi Source Catalog (2FGL, \citealt{Nolan12}). Radio pulsar searches targeting unidentified 2FGL sources with pulsar-like gamma-ray spectra have shown that MSPs are common gamma-ray emitters (e.g. \citealt{Ray12}). However, none of the pulsars in Table~\ref{tab_new} have 2FGL counterparts. Conversely, once we obtain timing solutions from radio timing observations, we plan to use them to fold Fermi event data and check whether any of the AO327 pulsars exhibit gamma-ray pulsations, which may be too weak for the objects to have been included in 2FGL. The ratio of radio-loud to radio-quiet gamma-ray pulsars is an especially good discriminator between gamma-ray pulsar population models, and radio surveys and multi-wavelength follow-up studies of new pulsars will make such distinction possible. 

The Arecibo Observatory is operated by SRI International under a cooperative agreement with the National Science Foundation (AST-1100968), and in alliance with Ana G. M\'{e}ndez-Universidad Metropolitana, and the Universities Space Research Association. M.A.M. is supported by the Research Corporation for Scientific Advancement. P.C.C.F. gratefully acknowledges financial support by the European Research Council for the ERC Starting Grant BEACON under contract 279702.

\end{document}